\documentstyle[aps,pra]{revtex}

\begin{document}
\newcommand\beq{\begin{equation}}
\newcommand\eeq{\end{equation}}
\newcommand\bea{\begin{eqnarray}}
\newcommand\eea{\end{eqnarray}}

\def\eps{\epsilon}
\newcommand{\ket}[1]{| #1 \rangle}
\newcommand{\bra}[1]{\langle #1 |}
\newcommand{\braket}[2]{\langle #1 | #2 \rangle}
\newcommand{\proj}[1]{| #1\rangle\!\langle #1 |}
\newcommand{\ba}{\begin{array}}
\newcommand{\ea}{\end{array}}
\newtheorem{theo}{Theorem}
\newtheorem{defi}{Definition}
\newtheorem{lem}{Lemma}
\newtheorem{exam}{Example}
\newtheorem{prop}{Property}
\newtheorem{propo}{Proposition}
\newtheorem{cor}{Corollary}
\newtheorem{conj}{Conjecture}


\author{Patrick Hayden$^1$, Barbara M. Terhal$^2$ and Armin Uhlmann$^3$}

\title{On the LOCC Classification of Bipartite Density Matrices}

\address{\vspace*{1.2ex}
	\hspace*{0.5ex}{$^1$Centre for Quantum Computation, 
	Clarendon Laboratory, Parks Road, Oxford, OX1 3PU, UK;\,}\\
	\hspace*{0.5ex}{$^2$IBM Watson Research Center,
	P.O. Box 218, Yorktown Heights, NY 10598, US;\,}\\
	\hspace*{0.5ex}{$^3$Institut f. Theoretische Physik,
	Universtit\"{a}t Leipzig, Augustusplatz 10/11, D-0409, Leipzig;\,}\\
	Emails: {\tt 
	patrick.hayden@qubit.org, 
	terhal@watson.ibm.com, 
	armin.uhlmann@itp.uni-leipzig.de}}

\date{\today}

\maketitle
\begin{abstract}
We provide a unifying framework for 
exact, probabilistic, and approximate conversions by local operations 
and classical communication (LOCC) between bipartite 
states. This framework
allows us to formulate necessary and sufficient conditions for LOCC 
conversions from pure states to mixed states and it 
provides necessary conditions for LOCC conversions between mixed states.
The central idea is the introduction of convex sets for exact, probabilistic, 
and approximate conversions, which are closed under LOCC operations and which are largely characterized by simple properties of pure states. 
\end{abstract}


\section{Introduction}

Bipartite entanglement has been shown to be a valuable resource in 
quantum information theory, allowing two parties, Alice and Bob to 
teleport unknown quantum states \cite{tele}, to remotely prepare 
quantum states \cite{lo,rsp} using classical communication only or 
to carry out other quantum information processing tasks. It is
therefore natural that a theory of bipartite entanglement has been developed 
which attempts to quantify this resource. In this theory, a central role
is played by the set of quantum operations which are constructed from 
local quantum operations by the parties Alice and Bob and classical 
communication between Alice and Bob, since these operations cannot enhance
the quantum correlations in a bipartite state. 
Here we will call these quantum operations LOCC operations, for
Local Operations and Classical Communication.
The goal of the theory of bipartite entanglement 
is to obtain a classification of states with respect to the set of 
LOCC operations.

Since the notion of majorization \cite{olkin} is to play a central role in
what follows, let us recall some helpful notation.
Let $\vec{\lambda}$ and $\vec{\mu}$ be two $n$-dimensional vectors 
with real coefficients in decreasing order, 
$\lambda^1 \geq \lambda^2 \geq \ldots \geq \lambda^n$, $\sum_{i=1}^n \lambda^i =1$ and likewise for $\vec{\mu}$. We write 
\bea
\vec{\lambda} \succ \vec{\mu} \Leftrightarrow \forall\, k=1,\ldots,n\, \sum_{i=1}^k \lambda^i \geq \sum_{i=1}^k \mu^i. 
\eea
When for all $l$ we have 
\beq
p \sum_{i=l}^n \lambda^{i} \leq \sum_{i=l}^n \mu^i,
\eeq
we say that $p \vec{\lambda} \succ^w \vec{\mu}$, or in words, $\vec{\mu}$ is 
weakly supermajorized by $p \vec{\lambda}$. For a bipartite pure state 
$\ket{\psi}$ we write $\vec{\lambda}_\psi$ for the vector
of Schmidt coefficients of $\ket{\psi}$ \emph{in decreasing order}.  (If
$\ket{\psi}$ is unnormalized then $\vec{\lambda}_\psi$ is associated with
the eigenvalues of $( {\rm Tr}_A \proj{\psi} ) / \braket{\psi}{\psi}$,
again in decreasing order.) 
In this paper we 
will call $|\langle \psi_1 | \psi_2 \rangle|$ and $F(\rho,\rho') = {\rm Tr}(\sqrt{\rho^{1/2}\rho'\rho^{1/2}})$ the square-root-fidelity between 
states $\ket{\psi_1}$ and $\ket{\psi_2}$ and $\rho$ and $\rho'$ respectively. 

Let us begin our study by recapturing what is known for LOCC conversions between 
pure states. Let $\ket{\psi_{\vec{\mu}}}$ be a state that we start with, 
which is characterized by a Schmidt vector $\vec{\mu}$.
\begin{enumerate}
\item Nielsen \cite{nielsen:major}: A pure state $\ket{\psi}$ can be obtained by LOCC from 
the state $\ket{\psi_{\vec{\mu}}}$ {\em exactly} if and only if 
\beq
\vec{\lambda}_{\psi} \succ \vec{\mu}.
\label{majcond}
\eeq
\item Vidal \cite{probvidal}: A pure state $\ket{\psi}$ can be obtained by LOCC from 
the state $\ket{\psi_{\vec{\mu}}}$ {\em with probability} $p$ if and only if 
\beq
p \vec{\lambda}_{\psi} \succ^w \vec{\mu}.
\label{condpurep}
\eeq
The protocol that achieves the optimal probability can be viewed as a 
two-step protocol 
\beq
\ket{\psi_{\vec{\mu}}} \rightarrow_{exact} \ket{\xi} \rightarrow_{{\cal M}_A} \ket{\psi},
\eeq
where $exact$ and ${\cal M}_A$ denote an exact conversion and a unilateral measurement by Alice respectively. The state $\ket{\xi}$ is an intermediate state 
which can be determined by a procedure given in Ref. \cite{probvidal}.
\item Vidal-Jonathan-Nielsen \cite{vjn}: The optimal $f=|\langle \psi|\psi' \rangle|$ with which we approximate a state $\ket{\psi}$ by a state $\ket{\psi'}$ which we obtain by LOCC from $\ket{\psi_{\vec{\mu}}}$ is equal to 
$|\langle \psi| \xi \rangle|$ where $\ket{\xi}$ is the intermediate state 
in the optimal probabilistic conversion of $\ket{\psi_{\vec{\mu}}}$ into 
$\ket{\psi}$.  (Note that in cases where the optimal probability is zero, the
protocol still allows for the construction of the state $\ket{\xi}$.)
\end{enumerate}

Our goal in this paper is to extend these results to the domain of 
mixed states. More specifically, we are interested in the `formation' 
problem for bipartite mixed states: Given a pure state 
$\ket{\psi_{\vec{\mu}}}$,
when can we obtain a density matrix $\rho$ from $\ket{\psi_{\vec{\mu}}}$ by 
LOCC, {\em exactly} or {\em with probability} $p$? Similarly, what is the 
maximum fidelity we can 
achieve when {\em approximating} $\rho$? The problem of entanglement 
distillation \cite{bdsw} for a mixed states is not covered by this 
formalism. What we will find is that the `formation' problem of mixed states 
can be completely translated to what is known for pure states, 
\emph{i.e.} the 3 items listed above. 
Our results are expressed in four theorems that we 
present in this section and that we will prove later on in the paper. 

Let us first define a restricted class of LOCC operations. We will call 
a quantum operation an ${\rm LOCC_1^A}$ operation when it is given by a local 
measurement/superoperator by Alice consisting of a set of commuting Kraus operators, followed by 1-way classical communication from 
Alice to Bob, and additional local unitary transformations by Alice and Bob, 
which may depend on Alice's measurement outcomes. Similarly, the class
of ${\rm LOCC_1^B}$ maps is the set of ${\rm LOCC_1^A}$ maps with Alice and Bob
interchanged.

For every pure state $\ket{\psi_{\vec{\mu}}}$ we define three kinds of 
sets, which we call $S_{\vec{\mu}}^{+}$, $S_{\vec{\mu},p}^{+}$ and $S_{\vec{\mu},f}^{+}$, 
where the first set relates to exact conversions, the second to probabilistic
conversions and the last one to approximate conversions. These sets are defined in the following way

\begin{defi}[Exact]
The set $S_{\vec{\mu}}$ is the convex closure of the set of pure states
$\ket{\psi}$ such that 
\beq
\vec{\lambda}_{\psi} \succ \vec{\mu}.
\eeq
The set $S_{\vec{\mu}}^{+}$ is defined as 
\beq
S_{\vec{\mu}}^{+}=\{{\cal L}(\ket{\psi}\bra{\psi})\,|\, {\cal L} \in {\rm LOCC_1^A},\; \ket{\psi} \in S_{\vec{\mu}} \}.
\eeq
\label{defiexact}
\end{defi}

Note that the sets $S_{\vec{\mu}}$ are generalizations of the Schmidt number sets $S_k$
which were introduced in Ref. \cite{terhalsrank}. A Schmidt number set $S_k$
is a set $S_{\vec{\mu}}$ where $\vec{\mu}$ are the Schmidt coefficients
of a maximally entangled state of Schmidt rank $k$.

\begin{defi}[Probabilistic]
The set $S_{\vec{\mu},p}$ is the convex closure of the set of pure states
$\ket{\psi}$ such that
\beq
p \vec{\lambda}_{\psi} \succ^w \vec{\mu}.
\eeq
The set $S_{\vec{\mu},p}^{+}$ is defined as 
\beq
S_{\vec{\mu},p}^{+}=\{{\cal L}(\ket{\psi}\bra{\psi})\,|\,  
{\cal L} \in {\rm LOCC_1^A},\; \ket{\psi} \in S_{\vec{\mu},p} \}.
\eeq
\label{defiprob}
\end{defi}

In Section \ref{krelate} we will discuss the relation between the sets $S_{\vec{\mu},p}^{+}$ and $S_k$.

\begin{defi}[Approximate]
The set $S_{\vec{\mu},f}$ is the convex closure of the set of pure states
$\ket{\psi}$ such that there exists a pure state $\ket{\xi} \in S_{\vec{\mu}}$ 
with $|\langle \xi| \psi\rangle| \geq f$.
The set $S_{\vec{\mu},f}^{+}$ is defined as 
\beq
S_{\vec{\mu},f}^{+}=\{{\cal L}(\ket{\psi}\bra{\psi})\,|\, 
{\cal L} \in {\rm LOCC_1^A},\; \ket{\psi} \in S_{\vec{\mu},f} \}.
\eeq
\label{defiapprox}
\end{defi}

It is not hard to see that the sets given in the Definitions \ref{defiexact}-\ref{defiapprox} are all compact, convex sets. In Appendix \ref{posmap} we will 
prove that unlike the Schmidt number sets, these sets do not relate directly
to positive linear maps. Nonetheless, 
the characterization of these sets in terms of a pure state and an 
${\rm LOCC_1^A}$ operation is extremely simple. We are able to prove the following three theorems.  The first is essentially a re-statement
of a result by Jonathan and Plenio \cite{jonaplenio}
in the language of $S_{\vec{\mu}}^{+}$ sets.

\begin{theo}
A density matrix $\rho \in S_{\vec{\mu}}^{+}$ if and only if there exists a 
decomposition $\{p_i,\ket{\psi_i}\}$ of $\rho$ such that 
\beq
\sum_i p_i \vec{\lambda}_{\psi_i} \succ \vec{\mu}.
\eeq
Furthermore, the LOCC conversion 
\beq
\ket{\psi_{\vec{\mu}}} \rightarrow_{exact} \rho,
\eeq
is possible if and only if $\rho \in S_{\vec{\mu}}^{+}$.
\label{theoexact}
\end{theo}

A straightforward corollary of this theorem is 

\begin{cor}
Let $\rho \in S_{\vec{\mu}}^{+}$, then $E(\rho) \leq H(\vec{\mu})$ where 
$H(\vec{\mu})$ is the Shannon entropy of the vector $\vec{\mu}$ and $E$ is 
the entanglement of formation of $\rho$,
\beq
E(\rho)=\min_{{\cal E}=\{p_i,\ket{\psi_i}:\rho=\sum_i p_i \ket{\psi_i}\bra{\psi_i}\}} \sum_i p_i E(\ket{\psi_i}\bra{\psi_i}),
\eeq 
where $E(\proj{\psi})$ is the von Neumann entropy of the reduced density
matrix of $\ket{\psi}$.
\end{cor}

The second theorem, dealing with the case of probabilistic conversions
from pure to mixed states, combines the results in Ref. \cite{jonaplenio} on exact
conversions for mixed states with those of Vidal for probabilistic
conversions between pure states \cite{probvidal}.

\begin{theo}
A density matrix $\rho \in S_{\vec{\mu},p}^{+}$ if and only if there exists a 
decomposition $\{p_i,\ket{\psi_i}\}$ of $\rho$ such that 
\beq
p \sum_i p_i \vec{\lambda}_{\psi_i} \succ^w \vec{\mu}.
\eeq
Furthermore, the LOCC conversion 
\beq
\ket{\psi_{\vec{\mu}}} \rightarrow \rho,
\eeq
is possible with probability at least $p$ if and only if $\rho \in S_{\vec{\mu},p}^{+}$.
\label{theoprob}
\end{theo}

For the last theorem, we need to introduce the notion of 
$(\vec{\mu},f)$-approximability.

\begin{defi}A density matrix $\rho$ is $(\vec{\mu},f)$-approximable if and 
only if 
\beq
\sum_i |\langle \psi_i |\psi_i' \rangle| \geq f,
\eeq
for some decomposition $\rho=\sum_i \ket{\psi_i}\bra{\psi_i}$,
and set of states $\{\ket{\psi_i'}\}$ such that
${\rm Tr} \sum_i \proj{\psi_i'} = 1$,
$p_i=\langle \psi_i'|\psi_i' \rangle$, and
\beq
\sum_i p_i \vec{\lambda}_{\psi_i'} \succ \vec{\mu}.
\label{majens}
\eeq
\label{defapprox}
\end{defi}

{\em Remarks}
We could have associated a density matrix 
$\rho' = \sum_i \proj{\psi_i'} \in S_{\vec{\mu}}^{+}$ with the
set of approximating states $\{\ket{\psi_i'}\}$.
It is not known whether our definition of $(\vec{\mu},f)$-approximability 
is equivalent with the more natural definition in terms of 
the square root of the transition probability 
$F(\rho,\rho') = {\rm Tr}(\sqrt{\rho^{1/2}\rho'\rho^{1/2}})$; we could have 
called a state $\rho$ $(\vec{\mu},f)$-approximable iff there 
exists a $\rho' \in S_{\vec{\mu}}^{+}$ such that 
$F(\rho,\rho') \geq f$. Because of the strong concavity property of the 
square-root-fidelity \cite{uhlmann_fid},
\beq
F(\rho,\rho') 
\geq \sum_i | \braket{\psi_i}{\psi_i'} |
\label{eqnfidebound}
\eeq
whenever $\rho$ is $(\vec{\mu},f)$-approximable according to 
Definition \ref{defapprox}, it follows that $\rho$ is 
also approximable according to a definition which uses the square-root-fidelity.
Furthermore, even though there exists a pair of decompositions which 
saturate Eq. (\ref{eqnfidebound}), see Ref. \cite{uhlmann_new}, these optimal
decompositions need not correspond to ones which obey the relation of 
Eq. (\ref{majens}).  While we will further investigate the connection
between $F$ and $(\vec{\mu},f)$-approximability in Section \ref{secapprox}, it
remains unclear whether this second definition would allow for Theorem 3. 

The following theorem and its corollaries, when combined with
the Vidal-Jonathan-Nielsen results \cite{vjn}, 
create a framework for answering the question of approximating
a mixed state using LOCC operations, starting from a given pure state.

\begin{theo}
A density matrix $\rho \in S_{\vec{\mu},f}^{+}$ if and only if $\rho$ is 
$(\vec{\mu},f)$-approximable.  
\label{theoapprox}
\end{theo}

It turns out that for two qubits a simplification takes place:
\begin{theo}
For two qubits $S_{\vec{\mu},x}^{+}=S_{\vec{\mu},x}$ where $x$ refers to either 
exact, probabilistic or approximate conversions. Let $S$ be the set of all
bipartite 2-qubit density matrices. We have 
\beq
0 < q \leq 2 \mu_2:\;\; S_{(\mu_1,\mu_2),p=q}=S,
\eeq
and 
\beq
1 \geq q > 2 \mu_2:\;\; S_{(\mu_1,\mu_2),p=q}=S_{(\mu_2/q,1-\mu_2/q))}.
\eeq
\label{qubits}
\end{theo}

Thus for two qubits, the probabilistic sets are identical
to each other and identical to the exact sets.


Theorems \ref{theoexact}-\ref{theoapprox} also give rise to {\em necessary}
conditions for LOCC conversions from a mixed state to either a pure or 
mixed state.  From Theorem \ref{theoexact} we can conclude that 

\begin{cor}
The LOCC conversion
\beq
\rho_1 \rightarrow_{exact} \rho_2,
\eeq 
where $\rho_1$ is a mixed state is not possible if there exists a pure state 
$\ket{\psi_{\vec{\mu}}}$ such that $\rho_1 \in S_{\vec{\mu}}^{+}$, but $\rho_2 \not \in 
S_{\vec{\mu}}^{+}$.
\end{cor}

For probabilistic conversions, Theorem \ref{theoprob} leads immediately to 

\begin{cor}
Suppose that $\rho_1 \in S_{\vec{\mu},p}^{+}$ and that, in particular,
it is possible to obtain $\rho_1$ from $\ket{\psi_{\vec{\mu}}}$ with
some probability $p_1 \geq p$.  Now suppose $\rho_2 \not\in S_{\vec{\mu},p}^{+}$.
The maximum probability of success $p_2$ of obtaining $\rho_2$ from
$\rho_1$ via LOCC operations satisfies $p_2 < p / p_1$.
\end{cor}


\section{LOCC conversions}

\subsection{Preliminaries}

In proving Theorems \ref{theoexact} and \ref{theoprob} one observation 
turns out to be crucial.

\begin{quote}{\em Observation 1}
When a density matrix $\rho$ can be obtained by LOCC from a pure state 
$\ket{\psi_{\vec{\mu}}}$ with probability 
$p \in (0,1]$ then there exists some decomposition $\{p_i,\ket{\psi_i}\}$ of $\rho$ into pure states, \emph{i.e.} $\rho=\sum_i p_i \ket{\psi_i}\bra{\psi_i}$, such that we can obtain the ensemble of states 
$\ket{\psi_i}$ each with probability $p p_i$ by LOCC from the state $\ket{\psi_{\vec{\mu}}}$.
\end{quote} 

It is essential that the LOCC protocol starts with a pure state. Consider a 
unilateral operation by, say, Alice, on a pure state $\ket{\psi}$:
\beq
\ket{\psi} \rightarrow \{p_i, {\cal S}_i(\ket{\psi}\bra{\psi})\},
\eeq
where ${\cal S}_i$ is some CP map, characterized by Kraus operators 
$\{A_{k}^i\}$. Instead of performing ${\cal S}_i$, she can perform 
the more `fine-grained' measurement 
\beq
\ket{\psi} \rightarrow \left\{p_i q_{ij}, 
\frac{\ket{\psi_{ij}}\bra{\psi_{ij}}}{q_{ij}}\right\},
\eeq
where $\ket{\psi_{ij}}=A_j^i \otimes {\bf 1} \ket{\psi}$ and 
$q_{ij}={\rm Tr}\, {A_j^i}^{\dagger} A_j^i \otimes {\bf 1} \ket{\psi}\bra{\psi}$, which gives her an ensemble of pure states. The extra bits of 
information that she obtains will be communicated to Bob in the next round.
By replacing every `mixing' operator in the LOCC protocol by a fine-grained measurement projecting on pure states, Alice and Bob obtain a new protocol which produces 
an ensemble of pure states as output, and the ensemble forms a decomposition 
of the mixed state that was obtained in the original LOCC protocol. 

\subsection{Exact}

\subsubsection{Proof of Theorem \protect \ref{theoexact}}

The proof of this Theorem follows quite straightforwardly from 
Observation 1 and the results of Jonathan and Plenio \cite{jonaplenio}.
Theorem 1 of Ref. \cite{jonaplenio} states that we can obtain the ensemble
$\{p_i,\ket{\psi_i}\}$ ($\sum_i p_i=1$) by LOCC from a state $\ket{\psi_{\vec{\mu}}}$ if and only if 
\beq
\sum_i p_i \vec{\lambda}_{\psi_i} \succ \vec{\mu}.
\label{avmaj}
\eeq
Now, if we can obtain the ensemble $\{p_i,\ket{\psi_i}\}$ by LOCC, then 
we can obtain $\rho=\sum_i p_i \ket{\psi_i}\bra{\psi_i}$. On the other hand,
if we can obtain $\rho$ by LOCC from the state $\ket{\psi_{\vec{\mu}}}$, then 
we can obtain {\em some} decomposition of $\rho$ by Observation 1, and thus 
for this ensemble Eq. (\ref{avmaj}) should hold. 

The equivalence between Eq. (\ref{avmaj}) for some ensemble $\{p_i,\ket{\psi_i}\}$ of $\rho$ and $\rho \in S_{\vec{\mu}}^{+}$ follows from inspecting the Jonathan-Plenio \cite{jonaplenio} protocol for converting $\ket{\psi_{\vec{\mu}}}$ to the ensemble $\{p_i,\ket{\psi_i}\}$. It is a two step protocol:
\beq
\ket{\psi_{\vec{\mu}}} \rightarrow_{exact} \ket{\bar{\psi}} \rightarrow_{{\rm LOCC_1^A}} \rho,
\label{plenjon}
\eeq  
where $\ket{\bar{\psi}} \in S_{\vec{\mu}}$ and its Schmidt vector $\vec{\lambda}_{\bar{\psi}}$ is given 
by $\lambda_{\bar{\psi}}^i=\sum_j p_j \lambda_{\psi_j}^i$.
Thus, every $\rho$ for which there exists an ensemble obeying 
Eq. (\ref{avmaj}), can be written as 
$\rho={\cal L}(\ket{\psi}\bra{\psi})$ where ${\cal L}$ 
is some ${\rm LOCC_1^A}$ map and $\ket{\psi} \in S_{\vec{\mu}}$.
These observations prove Theorem \ref{theoexact}.
$\Box$

\subsubsection{Consequences}
 
Note that since the majorization conditions treat Alice and Bob on an 
equal basis, we can alternatively write
\beq
S_{\vec{\mu}}^{+}=\{{\cal L}(\ket{\psi}\bra{\psi})\,|\, {\cal L} \in {\rm LOCC_1^B},\, \ket{\psi} \in S_{\vec{\mu}} \}.
\eeq

One might ask whether the extra complication incurred by defining the
sets $S_{\vec{\mu}}^{+}$ is necessary.  While we will find later on that
in the case of pairs of qubits the sets $S_{\vec{\mu}}^{+}$ are equal
to the simpler sets $S_{\vec{\mu}}$, the following example illustrates
that for pairs of $n$-level systems, with $n \geq 3$, that the sets generally
do not coincide.
\begin{exam}
Let $\ket{\psi_0} = \ket{00}$ and 
$\ket{\psi_{12}} = \frac{1}{\sqrt{2}}(\ket{11}+\ket{22})$ then
consider the density operator
$\rho = (1-\epsilon)\proj{\psi_0} + \epsilon\proj{\psi_{12}}$.  Let
$\vec{\mu} = (1-\epsilon/2,\epsilon/2,0)$.  Then for $\epsilon \in (0,1)$,
$\rho \in S_{\vec{\mu}}^{+}$ but $\rho \not\in S_{\vec{\mu}}$.
\end{exam}
First, notice that $\vec{\lambda}_{\psi_0} = (1,0,0)$ while 
$\vec{\lambda}_{\psi_{12}} = (\frac{1}{2},\frac{1}{2},0)$.  Thus, 
$(1-\epsilon)\vec{\lambda}_{\psi_0}+\epsilon\vec{\lambda}_{\psi_{12}} =\vec{\mu}$
and we have that $\rho \in S_{\vec{\mu}}^{+}$.  

Now we need to demonstrate that for any decomposition
$\rho=\sum_i p_i \proj{\phi_i}$ there always exists an $i$ such
that $\vec{\lambda}_{\phi_i} \not\succ \vec{\mu}$.  Let 
\begin{equation}
P_{12} = (\proj{1} + \proj{2})\otimes (\proj{1} + \proj{2}),
\end{equation}
and $P_0 = \ket{00}\bra{00}$.  Then 
$P_{12} \rho P_{12} = \eps \proj{\psi_{12}}$ from which it follows that
$P_{12} \proj{\phi_i} P_{12} \propto \proj{\psi_{12}}$   
and similarly $P_0 \proj{\phi_i} P_0 \propto \proj{00}$. Together this implies that 
\beq
\ket{\phi_i} 
= (P_0 + P_{12})\ket{\phi_i}
= \alpha_i \ket{00} + \frac{\beta_i}{\sqrt{2}} \ket{11} + \frac{\beta_i}{\sqrt{2}} \ket{22},
\eeq
for some complex amplitudes $\alpha_i,\beta_i$.
Letting $\alpha_i^*\alpha_i = r_i$ and $\beta_i^*\beta_i = s_i$, we write
$\vec{\lambda}_{\phi_i} = (r_i,s_i/2,s_i/2)$, assuming $r_i > s_i/2$.  If both 
$r_i$ and $s_i$ are nonzero, then $\vec{\lambda}_{\phi_i} \not \succ \vec{\mu}$.
On the other hand, any decomposition of
$\rho$ into pure states each with either $r_i=0$ or $s_i=0$
will necessarily contain terms for which $\ket{\phi_i}=\ket{\psi_{12}}$.
Thus, $\rho \not\in S_{\vec{\mu}}$.  $\Box$


\subsection{Probabilistic}

\subsubsection{Proof of Theorem \protect \ref{theoprob}}

First, we prove that if a state $\rho$ has a decomposition 
$\rho=\sum_j p_j \ket{\psi_j}\bra{\psi_j}$
satisfying the condition
\beq
p \sum_j p_j \vec{\lambda}_{\psi_j} \succ^w \vec{\mu},
\label{avmajp}
\eeq
then $\rho \in S_{\vec{\mu},p}^{+}$. Let us define the state 
$\ket{\bar{\psi}}$ with Schmidt coefficients $\lambda^i=\sum_j p_j \lambda_{\psi_j}^i$. Eq. (\ref{avmajp}) directly implies that $\ket{\bar{\psi}} \in S_{\vec{\mu},p}$ and thus by Vidal's result on probabilistic pure state conversions,
Eq. (\ref{condpurep}), we can obtain $\ket{\bar{\psi}}$ with probability $p$ by LOCC from $\ket{\psi_{\vec{\mu}}}$. Then an ${\rm LOCC_1^A}$ map as in 
the last step of Eq. (\ref{plenjon}) results in $\rho$. 
This shows that that $\rho \in S_{\vec{\mu},p}^{+}$ (and also that we can obtain $\rho$ by LOCC with probability $p$).

On the other hand, let $\rho \in S_{\vec{\mu},p}^{+}$, 
\emph{i.e.} $\rho={\cal L}(\ket{\psi}\bra{\psi})$ with ${\cal L} \in {\rm LOCC_A^1}$ and 
 $\ket{\psi} \in S_{\vec{\mu},p}$. This implies that we can obtain $\rho$ 
with probability at least $p$ by LOCC from $\ket{\psi_{\vec{\mu}}}$.
By Theorem 1 we have that there exists a decomposition $\{p_i,\ket{\psi_i}\}$ 
of $\rho$ such that 
\beq
\sum_i p_i \vec{\lambda}_{\psi_i} \succ \vec{\lambda}_{\psi}, 
\eeq
and by Eq. (\ref{condpurep}), we get 
\beq
p \sum_i p_i \vec{\lambda}_{\psi_i} \succ p \vec{\lambda}_{\psi} \succ^w \vec{\mu}, 
\eeq
which is the desired condition.
$\Box$


As before, this result allows us to write down an
alternative characterization of the sets $S_{\vec{\mu},p}^{+}$, namely
\beq
S_{\vec{\mu},p}^{+}=\{{\cal L}(\ket{\psi}\bra{\psi})\,|\, {\cal L} \in 
{\rm LOCC_1^B},\, \ket{\psi} \in S_{\vec{\mu},p} \}.
\eeq

\subsubsection{Relation with Schmidt number sets}
\label{krelate}
 
What is the relation of the sets $S_{\vec{\mu},p}^{+}$ with the Schmidt 
number sets $S_k$?
For every vector $\vec{\mu}$, there is some probability $p$ by which 
$\ket{\psi_{\vec{\mu}}}$ can be mapped onto $\ket{\Psi_k}$ where $k$ 
is the Schmidt rank of $\vec{\mu}$. Therefore $S_k \subseteq S_{\vec{\mu},p_{max}}^{+}$ 
where $p_{max}$ is given by the maximum probability $p$ such that 
$p \vec{\lambda}_{\Psi_k} \succ^w \vec{\mu}$. Note also that  
$S_{\vec{\mu},p_{max}}^{+} \subseteq S_{\vec{\mu},q}^{+}$ where $q \leq  p_{max}$.
On the other hand, we have 
$S_{\vec{\mu},p}^{+} \subseteq S_k$ for all $p > 0$. This follows from 
the fact that (1) $\ket{\psi_{\vec{\mu}}}$ has Schmidt rank $k$ and (2) any state $\ket{\psi_{\mu}}$ can be reached by LOCC from $\ket{\Psi_k}$ with 
probability 1, due to the condition in Eq. (\ref{majcond}), and (3) from $\ket{\psi_{\vec{\mu}}}$ all states in $S_{\vec{\mu},p}^{+}$ can be obtained by LOCC with probability at least $p$. 

Therefore
\beq
S_k=S_{\vec{\mu},q}^{+}, \; \forall \, 0 < q \leq p_{max}, 
\eeq
that is, for small probability up to some maximum, the sets $S_{\vec{\mu},q}^{+}$ collapse onto the set $S_k$ and onto each other. 

\subsection{Approximate conversions}
\label{secapprox}

The following definition will be useful in what follows.
\begin{defi}[Optimal $\vec{\mu}$-approximation] A density matrix 
$\rho'=\sum_i \proj{\psi_i'}$, where $p_i=\braket{\psi_i'}{\psi_i'}$ and 
\beq
\sum_i p_i \vec{\lambda}_{\psi_i'} \succ \vec{\mu},
\label{condmajo}
\eeq
is an
\emph{optimal $\vec{\mu}$-approximation} to $\rho=\sum_i \ket{\psi_i}\bra{\psi_i}$ when
\beq
\sum_i |\langle \psi_i|\psi_i'\rangle|
\eeq
is maximal with respect to all decompositions of $\rho$ and 
sets of states $\ket{\psi_i'}$ for which condition Eq. (\ref{condmajo}) holds.
\label{defoptapprox}
\end{defi}

\subsubsection{Proof of Theorem \protect \ref{theoapprox}}

Let $\rho$ be a $(\vec{\mu},f)$-approximable state and $\rho'$ be
an optimal $\vec{\mu}$-approximation to $\rho$.  Thus there exist decompositions $\rho=\sum_i \ket{\psi_i} \bra{\psi_i}$ and 
$\rho'=\sum_i \ket{\psi'_i}\bra{\psi'_i}$
such that
\beq
\sum_i | \braket{\psi_i}{\psi'_i} | \geq f.
\eeq
Since $\rho' \in S_{\vec{\mu}}^{+}$, we can find unitary operators
$U_i'$ and $V_i'$, non-negative diagonal Kraus operators 
$A_i = {\rm diag}(\sqrt{\lambda_{i,1}},\ldots,\sqrt{\lambda_{i,n}})$ 
satisfying $\sum_i A_i^2 = {\bf 1}$
and a state $\ket{\psi_{\rho'}} \in S_{\vec{\mu}}$
such that $\ket{\psi'_i} = (U_i' \otimes V_i' A_i) \ket{\psi_{\rho'}}$.
Similarly, we can find unitary $U_i$ and $V_i$,
operators $B_i = {\rm diag}(\sqrt{\mu_{i,1}},\ldots,\sqrt{\mu_{i,n}})$, 
$\sum_i B_i^2 = {\bf 1}$ and a state 
$\ket{\psi_\rho}$ such that
$\ket{\psi_i} = (U_i \otimes V_i B_i) \ket{\psi_\rho}$.  
Furthermore, the Plenio-Jonathan protocol requires that 
$\ket{\psi_\rho}$ and $\ket{\psi_{\rho'}}$ be in Schmidt form
in the bases that diagonalize their respective Kraus operators
$A_i$ and $B_i$, which in both cases is the standard basis.
Therefore,
$\ket{\psi_{\rho'}} = \sum_k \sqrt{\alpha_k} \ket{k,k}$ and
$\ket{\psi_\rho} = \sum_k \sqrt{\beta_k} \ket{k,k}$.

By Lemma 1 of Ref. \cite{vjn}, the absolute value of the inner product
between two states $\ket{\psi}$ and $\ket{\psi'}$ with given
Schmidt coefficients is maximized when the states are chosen to have
the same Schmidt basis.  We can therefore assume that
$U_i' = U_i$ and $V_i' = V_i$ since $\rho'$ is an optimal
$\vec{\mu}$-approximation to $\rho$.  We can then calculate
\beq
\sum_i |\braket{\psi_i}{\psi_i'}| 
= \sum_{i,k} \sqrt{\alpha_k \lambda_{ik}} 
	\sqrt{\beta_k \lambda_{ik}}.
\eeq
However, since $\sum_i \lambda_{i,k} = \sum_i \mu_{i,k} = 1$, we know
that $\sum_i \sqrt{ \lambda_{i,k} \mu_{i,k} } \leq 1$.  Therefore,
again using the optimality of $\rho'$, we can conclude that
$\mu_{i,k} = \lambda_{i,k}$.  It follows that
\beq
|\braket{\psi_\rho}{\psi_{\rho'}}| 
= \sum_i | \braket{\psi_i}{\psi_i'} |
\geq f.
\eeq
Therefore $\ket{\psi_{\rho}} \in S_{\vec{\mu},f}$. We have shown that when 
$\rho$ is $(\vec{\mu},f)$-approximable, we can write 
$\rho={\cal L}(\ket{\psi_{\rho}}\bra{\psi_{\rho}})$ 
where $\ket{\psi_{\rho}} \in S_{\vec{\mu},f}$ and ${\cal L}$ is an 
${\rm LOCC_1^A}$ map.

On the other hand, let $\rho \in S_{\vec{\mu},f}^{+}$; we can write 
$\rho={\cal L}(\ket{\psi}\bra{\psi})$ where $\ket{\psi} \in S_{\vec{\mu},f}$. 
Consider an optimal $\vec{\mu}$-approximation 
$\ket{\psi'} \in S_{\vec{\mu}}$ to $\ket{\psi}$. As an approximation 
to $\rho$ we take 
$\rho'={\cal L}(\ket{\psi'}\bra{\psi'})$. We write 
$\rho=\sum_i \ket{\psi_i}\bra{\psi_i}$ and 
$\rho'=\sum_i \ket{\psi'_i}\bra{\psi'_i}$ where 
$(U_i \otimes A_i) \ket{\psi}=\ket{\psi_i}$ and 
similarly for $\ket{\psi'_i}$, where $A_i$ are the Kraus operators of 
a local superoperator implemented by Alice. Then we have 
\beq
\sum_i |\langle \psi_i| \psi'_i \rangle|=
\sum_i |\langle \psi | A_i^{\dagger}A_i \otimes {\bf 1}|\psi' \rangle|
\geq |\langle \psi| \psi' \rangle| \geq f, 
\eeq
where we used $\sum_i |x_i| \geq |\sum_i x_i|$.  Therefore $\rho$ is $(\vec{\mu},f)$-approximable. $\Box$

\subsubsection{Connection to square-root fidelity}

We will now investigate in more detail the relationship between
square-root fidelity and $(\vec{\mu},f)$-approximability.  For
convenience, we begin with

\begin{defi}[$\vec{\mu}$-fidelity]
Let $\rho$ and $\sigma$ be density operators.  The $\vec{\mu}$-fidelity is
defined to be
$f_{\vec{\mu}}(\rho,\sigma) = \max \sum_i | \braket{\phi_i}{\psi_i} |$, where
$\rho = \sum_i \proj{\phi_i}$ and $\sigma = \sum_i \proj{\psi_i}$ with
$\sum_i p_i \vec{\lambda}_{\psi_i} \succ \vec{\mu}$ for 
$p_i = \braket{\psi_i}{\psi_i}$.  If no such decomposition 
exists then set $f_{\vec{\mu}}(\rho,\sigma) = 0$.
\end{defi}

Furthermore, define the following two optimal fidelities:
\begin{eqnarray}
F_{max}(\rho,\vec{\mu}) 
&=& \max \{ F(\rho,\sigma) : \sigma \in S_{\vec{\mu}}^{+} \} \\
f_{max}(\rho,\vec{\mu}) 
&=& \max \{ f_{\vec{\mu}}(\rho,\sigma) : \sigma \in S_{\vec{\mu}}^{+} \}.
\end{eqnarray}
In this section, we will develop some properties of $F_{max}$ and $f_{max}$
that might shed some light on the relationship between them.
We can concisely state our results on approximation by noting that
$F_{max}(\rho,\vec{\mu}) \geq f_{max}(\rho,\vec{\mu})$ and
$\rho \in S_{\vec{\mu},f}^{+}$ if and only if 
$f_{max}(\rho,\vec{\mu}) \geq f$.  However, since $F_{max}$ is a more
natural definition of the optimal fidelity of approximation, we would
ideally like to re-phrase the second result in terms of $F_{max}$.  We
will not be able to do so but can offer some suggestive partial results.

To begin, both functions are easily seen to be consistent with the partial
ordering on Schmidt vectors: $\vec{\lambda} \succ \vec{\mu}$ implies that
$F_{max}(\rho,\vec{\lambda}) \leq F_{max}(\rho,\vec{\mu})$, and similarly for
$f_{max}$.  Furthermore, notice that for a pure state
$\ket{\phi}$ and Schmidt vector $\vec{\mu}$, 
\begin{equation}
f_{max}(\proj{\phi},\vec{\mu}) = F_{max}(\proj{\phi},\vec{\mu}).
\end{equation}
One way to see this is to recall the result from Ref. \cite{vjn}
that the best approximation with respect to $F$ to a given pure state 
$\ket{\phi}$ by a mixed state $\sigma \in S_{\vec{\mu}}^{+}$ can always 
itself be taken to be given by a pure state 
$\ket{\psi} \in S_{\vec{\mu}}^{+}$.  
The result then follows because $f_{\vec{\mu}}(\ket{\phi},\ket{\psi})$ 
and $F(\ket{\phi},\ket{\psi})$ agree if $\lambda_{\ket{\psi}} \succ \vec{\mu}$.

The functions are similar in other ways as well, both them obeying a
strong form of joint concavity.  Let us begin with $F_{max}$.
Suppose that 
$F_{max}(\rho_j,\vec{\mu}_j) = F(\rho_j,\sigma_j)$ for some 
$\sigma_j \in S_{\vec{\mu}_j}^{+}$.  If we set
$\vec{\mu} = \sum_j q_j \vec{\mu}_j$, then we find 
$\sum_j q_j \sigma_j \in S_{\vec{\mu}}^{+}$ so that
\begin{eqnarray}
F_{max}(\sum_j p_j \rho_j, \sum_k q_k \vec{\mu}_k)
&\geq& F(\sum_j p_j \rho_j, \sum_k q_k \sigma_k) \\
&\geq& \sum_j \sqrt{p_j q_j} F(\rho_j, \sigma_j) \\
&=& \sum_j \sqrt{p_j q_j} F_{max}(\rho_j, \vec{\mu}_j).
\end{eqnarray}

Let us now move on to the joint concavity of $f_{max}$,
for which the proof is similar.  Suppose that
$f_{max}(\rho_j,\vec{\mu}_j) = f_{\vec{\mu}_j}(\rho_j,\sigma_j)$, where
$\sigma_j \in S_{\vec{\mu}_j}^{+}$.  Likewise, assume that the
optimal decomposition is given by
$f_{\vec{\mu}_j}(\rho_j,\sigma_j) = \sum_i | \braket{\phi_{ij}}{\psi_{ij}} |$,
where $\sum_i r_{ij} \vec{\lambda}_{\psi_{ij}} \succ \vec{\mu}_j$ for
$r_{ij} = \braket{\psi_{ij}}{\psi_{ij}}$.  Then, since
$\sum_{ij} q_j r_{ij} \vec{\lambda}_{\psi_{ij}} \succ \sum_j q_j \vec{\mu}_j$, we find
\begin{equation}
f_{max}(\sum_j p_j \rho_j, \sum_k q_k \vec{\mu}_k)
\geq \sum_{ij} \sqrt{p_j q_j} | \braket{\phi_{ij}}{\psi_{ij}} |
= \sum_j \sqrt{p_j q_j} f_{\vec{\mu}_j}(\rho_j,\sigma_j)
= \sum_j \sqrt{p_j q_j} f_{max}(\rho_j,\mu_j),
\end{equation}
which is the desired inequality.

Summarizing then, for a function $g$ equal to $F_{max}$ or $f_{max}$, we have
\begin{enumerate}
\item (Consistency with partial order) 
	$\vec{\lambda} \succ \vec{\mu}$ implies 
	$g(\rho,\vec{\lambda}) \leq g(\rho,\vec{\mu})$.
\item (Agreement on pure states)
	For a pure state $\ket{\phi}$ and Schmidt vector 
	$\vec{\mu}$, $g(\proj{\phi},\vec{\mu})$
	is equal to the maximum over $\ket{\psi} \in S_{\vec{\mu}}$ of
	$|\braket{\phi}{\psi}|$.
\item (Joint concavity)
	If $\rho = \sum_i p_i \proj{\phi_i}$ and 
	$\vec{\mu} =\sum_j q_j \vec{\mu}_j$ then
	$g(\rho,\vec{\mu}) \geq \sum_i \sqrt{p_i q_i} g(\rho_i,\vec{\mu}_i)$.
\end{enumerate}

There is a difference, however. For $f_{max}$ we also know that given 
$\rho$ and
$\vec{\mu}$, we can find a decomposition $\rho = \sum_i p_i \proj{\phi_i}$ and
an ensemble $\{q_i,\ket{\psi_i}\}$ with 
$\sum_i q_i \vec{\lambda}_{\psi_i} \succ \vec{\mu}$ 
such that 
$f_{max}(\rho,\vec{\mu}) = \sum_i \sqrt{p_i q_i} | \braket{\phi_i}{\psi_i} |$.
Since 
$|\braket{\phi_i}{\psi_i}| \leq f_{max}(\proj{\phi_i},\vec{\lambda}_{\psi_i})$ 
by definition, summing inequalities gives
\begin{equation}
f_{max}(\rho,\vec{\mu}) \leq \sum_i \sqrt{p_i q_i} f_{max}(\rho_i,\vec{\lambda}_{\psi_i}).
\end{equation}
Combining with joint concavity then implies that
\begin{equation}
f_{max}(\rho,\vec{\mu}) = \sum_i \sqrt{p_i q_i} f_{max}(\proj{\phi_i},
	\vec{\lambda}_{\psi_i}).
\end{equation}
Thus, $f_{max}$ is the smallest function consistent with the three properties
listed above.

This is quite a curious situation.  The results of Ref. \cite{uhlmann_new}
immediately imply that the square-root fidelity $F(\rho,\sigma)$ (note that the arguments of $F$ are different than those of $F_{max}$) is the
smallest function $g$ satisfying the following different list of properties:
\begin{enumerate}
\item (Agreement on pure states) 
	For normalized $\ket{\phi}$ and $\ket{\psi}$, 
	$g(\proj{\phi},\proj{\psi}) = |\braket{\phi}{\psi}|$.
\item (Joint concavity)
	If $\rho = \sum_i p_i \rho_i$ and $\sigma = \sum_j q_j \sigma_j$ then
	$g(\rho,\sigma) \geq \sum_i \sqrt{p_i q_i} g(\rho_i,\sigma_i)$.
\end{enumerate}
Thus $F$ arises as the smallest function satisfying a set of conditions
which appear very similar to those conditions satisfied by $F_{max}$ and
$f_{max}$. It seems plausible, therefore, that $F_{max}$, which
is defined in terms of $F$, might itself be the smallest
function consistent with properties 1-3 of $F_{max}$ and $f_{max}$.
If that is the case, then $f_{max} = F_{max}$.

\subsubsection{Applications}
\label{approxapplications}

Theorem 3 does not tell us how to determine the maximum $\vec{\mu}$-fidelity 
with which we can approximate a density matrix $\rho$, given a starting state
$\ket{\psi_{\vec{\mu}}}$.  To answer this question will generally require
a further optimization over decompositions of $\rho$.
Let us consider an example.


\begin{exam} Let $\ket{\Psi_m} = 1/\sqrt{m} \sum_{i=1}^m \ket{i_A} \ket{i_B}$
be a maximally entangled bipartite state of Schmidt rank $m$ and let
$\vec{\lambda}_m$ be the Schmidt vector of $\ket{\Psi_m}$.  The
best approximation to a state $\rho$ achievable starting
from $\ket{\Psi_m}$ using LOCC operations will have an optimal approximation 
with 
\beq
f_{max}(\rho,\vec{\lambda}_m) =(1 - E_{m+1}(\rho))^{1/2}, 
\eeq
where $E_l(\rho)$ is one of
Vidal's entanglement monotones \cite{probvidal}, defined as
\beq
\label{eqnmonotone}
E_l(\rho)=\min_{{\cal E}} \sum_i p_i E_l(\ket{\psi_i}\bra{\psi_i}),
\eeq
in which $E_l(\ket{\psi_i}\bra{\psi_i})=\sum_{k=l}^n \lambda_{\psi_i}^k$ 
and ${\cal E}=\{p_i,\ket{\psi_i}\}$ is an ensemble realizing $\rho$. 
\end{exam}
In Ref. \cite{vjn}, it is demonstrated that if 
$\ket{\phi} = \sum_{i=1}^n \sqrt{\beta_i} \ket{i_A} \ket{i_B}$,
with $\beta_i \geq \beta_{i+1}$ then
the best approximation to $\ket{\phi}$ achievable using LOCC
operations and starting from the state
$\ket{\Psi_m}$
has square-root-fidelity $f_{max} = \left(\sum_{i=1}^m \beta_i\right)^{1/2}$.
Theorem 3 shows that the optimal approximation for mixed states
is that derived from some Jonathan-Plenio
precursor state of the form 
$\sum_i \sqrt{\sum_j p_j \lambda_{\psi_j}^i} \ket{i_A} \ket{i_B}$,
where $\rho = \sum_j p_j \proj{\psi_j}$. 
Therefore, the best approximation
to the mixed state $\rho$ has fidelity
\beq
f_{max}(\rho,\vec{\lambda}_m) 
=\max \left( \sum_j p_j \sum_{i=1}^m \lambda_{\psi_j}^i \right)^{1/2}
= \left( 1 - \min \sum_j p_j \sum_{i=m+1}^n 
	\lambda_{\psi_j}^i \right)^{1/2},
\eeq
where the optimizations are to be performed over all decompositions of $\rho$.  
The last line of the above equation, however, is simply
$(1-E_{m+1}(\rho))^{1/2}$.
$\Box$



\subsection{Two Qubits: Proof of Theorem \protect \ref{qubits}}

\subsubsection{Exact}

First we relate the set $S_{\vec{\mu}}$ to the entanglement of formation:

\begin{propo}
The entanglement of formation $E$ of a density matrix in ${\bf C}_2 \otimes {\bf C}_2$ is 
\beq
E(\rho)=\min_{\vec{\mu}\,|\,\rho \in S_{\vec{\mu}}} H_2(\vec{\mu}),
\eeq 
where $H_2(.)$ is the binary entropy function, \emph{i.e.} $H_2(x)=-x \log x-(1-x) 
\log (1-x)$.
\end{propo}

{\em Proof} The Wootters formula for the entanglement of formation \cite{woot}
tells us that in the optimal decomposition $\{p_i,\ket{\psi_i}\}$ of $\rho$, every state $\ket{\psi_i}$ has an equal amount of entanglement. Therefore if $\rho$ has entanglement $E(\rho)$, the density matrix $\rho$ must be contained in the set $S_{\vec{\mu}}$ where $H_2(\vec{\mu})=E(\rho)$. Conversely, if 
$\rho \in S_{\vec{\mu}}$ then $E(\rho) \leq H_2(\vec{\mu})$.$\Box$ 

From this Proposition it follows that $S_{\vec{\mu}}$ must be closed under 
LOCC and therefore $S_{\vec{\mu}}=S_{\vec{\mu}}^{+}$. Furthermore, for two qubits 
we know that the partial order on the pure states 
induced by majorization is a total order, which is characterized by 
a single parameter, the smallest eigenvalue $\mu_2$ of the reduced 
density matrix of the pure state. Therefore we can characterize the 
vector $\vec{\mu}$ with this single parameter $\mu_2$ and it follows that $S_{\mu_2} \subset S_{\mu'_2}$ when 
$\mu_2 < \mu'_2$. 

As a corollary of this result we obtain that we can convert $\ket{\psi_{\vec{\mu}}}$ to $\rho$ by LOCC if and only if $\rho \in S_{\vec{\mu}}$, which means that 
$E(\rho) \leq H(\vec{\mu})$. Therefore the minimal entanglement costs 
for preparing a single copy of $\rho$ exactly is $E(\rho)$. This result 
has been independently found by Vidal \cite{vidaleform}.

\subsubsection{Probabilistic}

Let us now consider the probabilistic sets and prove that for 2 qubits 
$S_{\vec{\mu},p}^{+}=S_{\vec{\mu},p}$. First we note that the relation 
with the Schmidt number sets gives us the following:
\beq
S=S_2=S_{(\mu_1,\mu_2),p=q},
\eeq
when $q \leq p_{max} = 2 \mu_2$. 
Here $S_2$ is the Schmidt number 2 set, which is 
identical to the set of all bipartite two-qubit density matrices, $S$.
Now let $q > 2 \mu_2$. A density matrix $\rho \in S_{\vec{\mu},q}^{+}$ if and only if there exists
a decomposition $\{p_i,\ket{\psi_i}\}$ of $\rho$ such that
\beq
q \sum_i p_i \lambda_{\psi_i}^2 \leq \mu_2,
\eeq
where $\lambda_{\psi_i}^2$ is the smallest Schmidt coefficients of 
$\ket{\psi_i}$. Since $\mu_2/q < 1/2$, this condition is identical to the 
requirement that $\rho \in S_{\vec{\mu}'=(\mu_2/q,1-\mu_2/q)}^{+}$. Since 
$S_{\vec{\mu}}^{+}=S_{\vec{\mu}}$ for two qubits, this condition is again 
identical to requirement that there exists a decomposition $\{p_i,\ket{\psi_i}\}$ of $\rho$ such that for all $i$
\beq
q \lambda_{\psi_i}^2 \leq \mu_2,
\eeq
which implies that $\rho \in S_{\vec{\mu},q}$. So we have shown that 
for $q > 2 \mu_2$, the exact sets $S_{(\mu_2/q,1-\mu_2/q)}$ and 
the probabilistic sets $S_{(\mu_1,\mu_2),q}$ are, in fact, identical. 

\subsubsection{Approximate}
Finally, let us move on to the approximation sets.  
We'll begin by considering pure states.  Suppose that 
$\ket{\psi} \not\in S_{\vec{\mu}}$ and let $\ket{\psi'} \in S_{\vec{\mu}}$ such that 
$|\braket{\psi}{\psi'}| = f_{max}$ is optimal.  The condition
that $\ket{\psi} \in S_{\vec{\mu},f}$ is that $f_{max} \geq f$. 
Suppose now that $\ket{\psi}$ has Schmidt coefficients 
$(\alpha,1-\alpha)$, where $\alpha \geq 1/2$.
It is clear that $f_{max}$ is a strictly increasing function of
$\alpha$ on the interval $[1/2,\mu_1]$, from which it follows
that $f_{max} \geq f$, or, equivalently, that $\ket{\psi} \in S_{\vec{\mu},f}$,
if and only if $\alpha \geq \mu_1'$ for
some $\mu_1' \in [1/2,\mu_1]$.  If we set $\mu_2' = 1-\mu_1'$,
it follows immediately that $S_{\vec{\mu},f} = S_{\vec{\mu}'}$ since the
extreme points of these convex sets are pure states.  It then
follows from the definitions that $S_{\vec{\mu},f}^{+} = S_{\vec{\mu}'}^{+}$.
We saw above, however, that $S_{\vec{\mu}'}^{+} = S_{\vec{\mu}'}$.  Thus,
$S_{\vec{\mu},f}^{+} = S_{\vec{\mu},f}$.

\section{Conclusion}

We have provided a unifying framework for exact, probabilistic, and
approximate LOCC conversions from pure states to mixed states. 
In each case we have found criteria defining exactly when these
conversions are possible, with the caveat that the criteria are always
expressed in terms of the existence of `optimal' decompositions of the target
mixed state having some easily verified property.
This work does not address the question 
of how to find these optimal decompositions of a density matrix $\rho$ 
for exact, probabilistic or approximate conversions of an initial
state $\ket{\psi_{\vec{\mu}}}$ into the state $\rho$. This problem 
could be as hard a determining 
optimal decompositions for the entanglement of formation, but may 
well be simpler.  In addition to resolving that question, with our framework 
established, a host of open
questions present themselves.  The most pressing is the question of
determining whether the function $f_{max}(\rho,\vec{\mu})$ we have been
studying here is or is not equal to the more natural $F_{max}(\rho,\vec{\mu})$.
Another question of interest 
is the relation between the sets $S_{\vec{\mu}}^{+}$ and the optimal 
decomposition of $\rho$ with respect to its entanglement of formation.

\section{Acknowledgments} We would like to thank David DiVincenzo 
for helpful discussions and Daniel Jonathan for pointing out an error in 
an earlier version of this manuscript. PMH is grateful to the 
Rhodes Trust and the EU QAIP project for support and to the quantum 
information group at IBM for their hospitality while much of this work 
was being performed. BMT acknowledges support of the ARO under contract number DAAG-55-98-C-0041.

\appendix

\section{Positive Linear Maps}
\label{posmap}

We may ask whether, say, the sets $S_{\vec{\mu}}$ or $S_{\vec{\mu}}^{+}$ are characterized by positive linear maps. Let us define a $\vec{\mu}$-positive linear map in the following way, as 
a generalization of $k$-positivity.
Here $\vec{\mu}$ is a $n$-dimensional Schmidt vector. We would propose the following definition.
The linear Hermiticity-preserving map ${\cal L}$ is positive with respect to 
$\ket{\psi_{\vec{\mu}}}$, or $\vec{\mu}$-positive if 
$({\bf 1} \otimes {\cal L})(\rho) \geq 0$ for all density matrices $\rho \in S_{\vec{\mu}}$. It turns out this implies that ${\cal L}$ is $k$-positive 
where $k$ is the Schmidt rank of the vector $\vec{\mu}$. The property of $\vec{\mu}$-positivity implies that for all
$\ket{\psi} \in S_{\vec{\mu}}$ and for arbitrary $\ket{\phi}$ we have
\beq
\bra{\phi}({\bf 1} \otimes {\cal L})(\ket{\psi}\bra{\psi}) \ket{\phi} \geq 0.
\eeq
We can always write $\ket{\phi}=(A \otimes {\bf 1})\ket{\Phi}$ where $\ket{\Phi}$ 
is some maximally entangled state. By commuting $A$ through, we obtain
\beq
\bra{\Phi}({\bf 1} \otimes {\cal L})(\ket{\psi_A}\bra{\psi_A}) \ket{\Phi} \geq 0,
\label{mukpos}
\eeq
where $\ket{\psi_A}=(A^{\dagger} \otimes {\bf 1})\ket{\psi}$. There is 
always an operator $A$ and therefore a state $\ket{\phi}$ and $\ket{\Phi}$ such that 
$\ket{\psi_A}$ is proportional to the maximally entangled state with 
Schmidt rank $k$, where $k$ is the Schmidt rank of $\ket{\psi}$ itself.
Then the condition in Eq. (\ref{mukpos}) becomes the condition for 
$k$-positivity of ${\cal L}$ as given in Ref. \cite{terhalsrank}. Thus $\vec{\mu}$-
positivity implies $k$-positivity, whereas $S_{\vec{\mu}} \subset S_k$.



\end{document}